\documentclass[usenatbib]{mnras} 
\usepackage{graphicx,graphics,amsmath}
\usepackage{float}

\usepackage{xcolor}

\usepackage[T1]{fontenc}

\onecolumn

\DeclareRobustCommand{\VAN}[3]{#2}
\let\VANthebibliography\thebibliography
\def\thebibliography{\DeclareRobustCommand{\VAN}[3]{##3}\VANthebibliography}

\title{Spin-down induced quark-hadron phase transition in cold isolated neutron stars}

\author[R. Prasad, \& Ritam Mallick ]
{R. Prasad$^{1}$\thanks{rprasad@iiserb.ac.in}, \& Ritam Mallick$^{1}$\thanks{mallick@iiserb.ac.in} \\
	$^{1}$ Indian Institute of Science Education and Research Bhopal, Bhopal -462066, India}
 
\date{Accepted XXX. Received YYY; in original form ZZZ}

\pubyear{2022}

\begin{document}
\label{firstpage}
\pagerange{\pageref{firstpage}--\pageref{lastpage}}

\maketitle

\begin{abstract}
We have studied the spin-down induced phase transition in cold, isolated neutron stars in this work. After birth, as the star slows down, its central density rises and crosses the critical density of phase transition, and a quark core is seeded inside the star. Intermediate mass stars are more likely to have a quark seeding in their lifetime at birth. Smaller neutron stars do not have a quark core and remain neutron stars throughout their life, whereas in massive stars, a quark core exists at their center from birth. In intermediate and massive stars, the quark core grows further as the star slows down. The appearance of a quark core leads to a sudden change in the moment of inertia of the star in its evolutionary history, and it is also reflected in a sudden discontinuity in the braking index of the star (at the frequency where the quark core first seeds). The energy released during the phase transition process as the quark core is seeded can excite the f-mode oscillation in the star and is emitted in the form of the gravitational wave, which is in the range of detection with present operating detectors; however, future detectors will enable a more clean extraction of this signals. Also, neutrinos and bursts of gamma-rays can originate from phase transition events. The spin-down induced phase transition could be gradual or in the form of subsequent leaps producing persistent or multiple transient emissions.
\end{abstract} 
 

\section{Introduction}
The existence and identification of quark stars (stars having quark cores, and outer nucleonic matter) is an open problem in astrophysics. The quest to find such stars has attained a lot of focus in recent years. This is an integral problem not only for astrophysicists but also for the nuclear and particle physicist because the density at the core of neutron stars (NSs) lies in the intermediate density range. At such densities, theoretical calculations are not possible (both ab-intio lattice and perturbative calculations), and even earth-based experiments have still not materialized \citep{Bandyopadhyay2017,2017IJMPD..2630015G}. The low-temperature intermediate density range is also interesting in the sense that under such conditions, a first-order phase transition (PT) from hadronic matter to quark matter is expected 
\citep{witten1984cosmic,itoh1970hydrostatic,alcock1986strange,PhysRevD.46.1274}.

The task of understanding the matter properties at NS cores is not easy due to the fact that they are not directly visible to us. Therefore, one has to model the NS from the core to the surface and then check their properties with signals we have through observation. In this direction, various theoretical works to model the equation of state (EoS), structural properties, possible constraints on mass, radius, tidal deformability, etc. have been carried out \citep{universe5070159, Alford_2019,  chatziioannou2020neutron, Lattimer2021review}. In the last $10-15$ years, various precise observations have been able to put severe constrain on the properties of NS \citep{Ozel:2006bv,2010Natur.467.1081D,PhysRevLett.119.161101, Riley_2019,Riley_2021}. Observation of massive pulsars has ruled out the property of soft EoS \citep{Greif_2020}, gravitational wave (GW) observation of GW170817 has put constrain on the tidal deformability, which also put constrain on the compactness of the NS, thereby also ruling out several EoS \citep{PhysRevLett.121.161101,PhysRevLett.120.172703,PhysRevLett.120.261103}. The recent results from NISER experiments have given a severe bound on the radius of a $1.4$ solar mass NS, thereby constraining the parameter space of EoS further \citep{Miller_2019,Miller_2021}. However, the existence of a quark core at the center of an NS is still a viable scenario, and some suggest it may be the most probable one \citep{annala2020evidence}.

The quark-hadron PT in the dense core of NSs is one of the formation channels which can result in the formation of quark stars (QSs)/ Hybrid Stars (HSs). The formation of quark core in NS interior may happen at the proto-neutron star (PNS) stage after core-collapse \citep{PhysRevLett.102.081101,Zha_2021}, in cold NSs in their lifetime \citep{PhysRevC.74.065804,PhysRevC.82.062801,PhysRevD.84.083002,Prasad_2018}, and the hypermassive neutron stars (HMNs) formed after the binary merger \citep{PhysRevLett.122.061101,PhysRevLett.122.061102}. The dynamical process leading to formation is different for different scenarios. In the case of PNSs, the collapsing core of the progenitor can attain a density a few times higher than the nuclear saturation density leading to a quark matter core. In cold NSs, spin changes or accretion can lead to the deconfinement and appearance of quark seed near the center, which may create a shock propagation or conversion front propagation from center to surface converting hadronic matter to quark matter up to some distance
\citep{PhysRevC.74.065804,PhysRevC.76.052801,Drago_2007,PhysRevC.82.062801,PhysRevD.84.083002}. In the binary merger case, both density rise driven conversion \citep{PhysRevLett.124.171103} or shock driven conversion are possible. There is, however, a fundamental difference between the PT scenarios of binary NS mergers and core-collapse with that of cold NSs. For the former case, the PT is at finite density and finite temperature ($20-80$ MeV), whereas for the latter, the PT is at finite density but almost a very low temperature ($1-10$ MeV). 

The PT process can lead to observational signatures in the form of neutrino emission \citep{BHATTACHARYYA2006195,PhysRevD.87.103007}, gamma-ray bursts (GRBs) \citep{Fryer_1998,Bombaci_2000,drago2004supernova,MALLICK201496}, GWs \citep{marranghello2002phase,PhysRevC.83.045802,Orsaria_2019} and even electromagnetic radiations \citep{Jaikumar:2006qx,PhysRevD.88.083006}. Observing these signals could provide evidence of QSs existence and will lead to the inclusion of a new star to the existing family of compact stars, and will also provide insight into the high-density matter. The phase-transition process time scale and gravitational wave signature obtained in our recent work for isolated NSs involve a two-step process, wherein hadronic matter to 2-flavor quark matter (u,d) conversion proceeds via deconfinement and 2-flavor quark matter (u,d) to 3-flavor quark matter (u,d,s) conversion via weak decay \citep{PhysRevC.74.065804}. The first step's GW signal frequency is about 100 kHz and lasts for 30-50 microseconds \citep{Prasad_2018, Prasad_2020}, and is out of reach for the presently operating GW detectors. In contrast, for the second step GW signals, the frequency is about a few 100 Hz, and the signal duration is ten milliseconds. For sources located below 100 Kpc distance, the PT process could leave imprints in currently operating LIGO and VIRGO detectors \citep{10.1093/mnras/stab2217}. However, these signals are burst type and short-lived in nature, and hard to extract from the detector data. Also, the PT process can leave its imprints on continuous GW emission and radio emission from pulsars \citep{10.1093/mnras/stab2217, PhysRevC.105.065807}. In the present work, we model the spin-down induced PT in the cold, isolated NSs and obtain other associated signatures that could help in identifying the PT process more systematically. 

\section{Hadronic and quark matter}
\begin{figure}
\includegraphics[height=2.8in,width=3.4in]{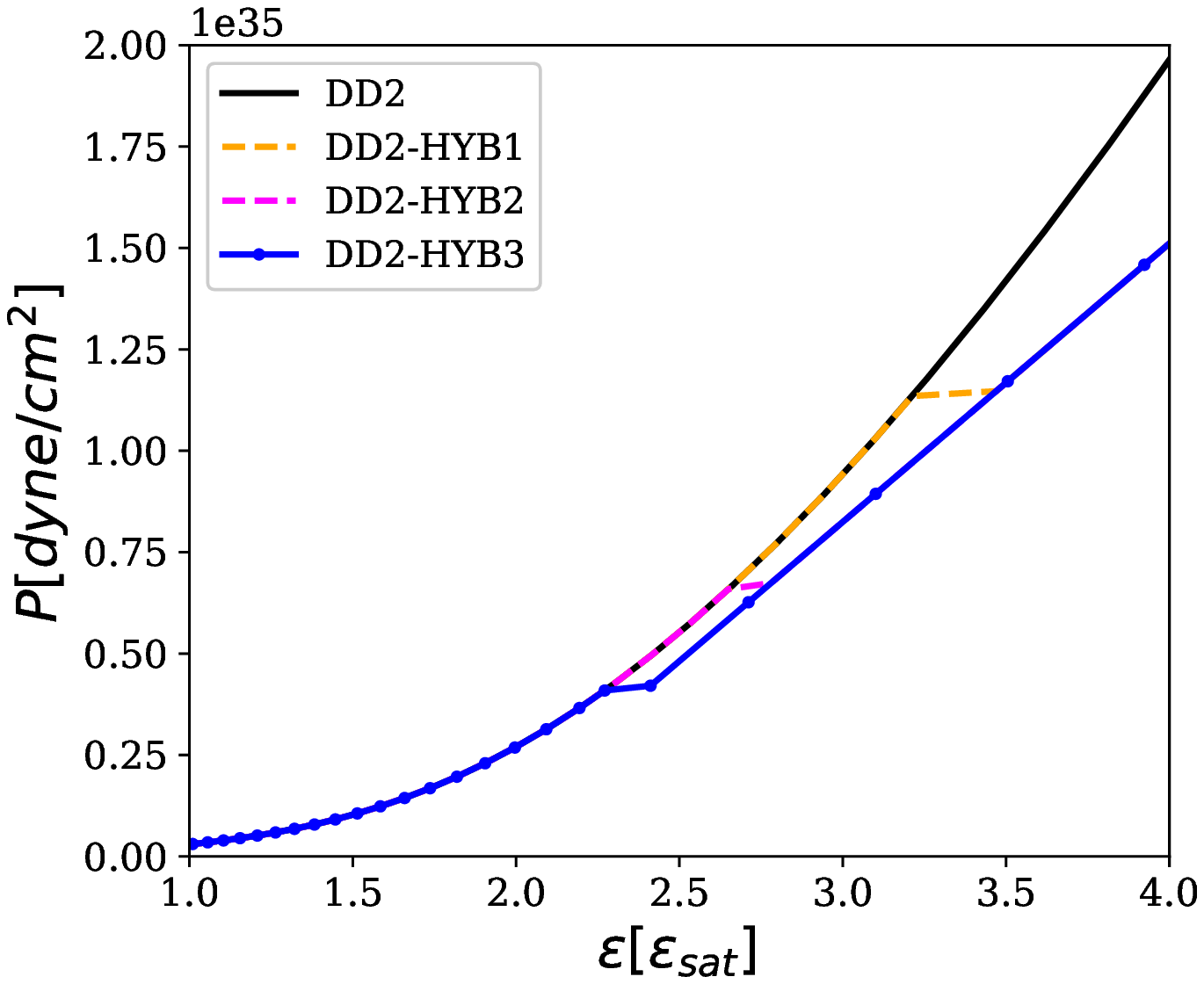}
\includegraphics[height=2.8in,width=3.4in]{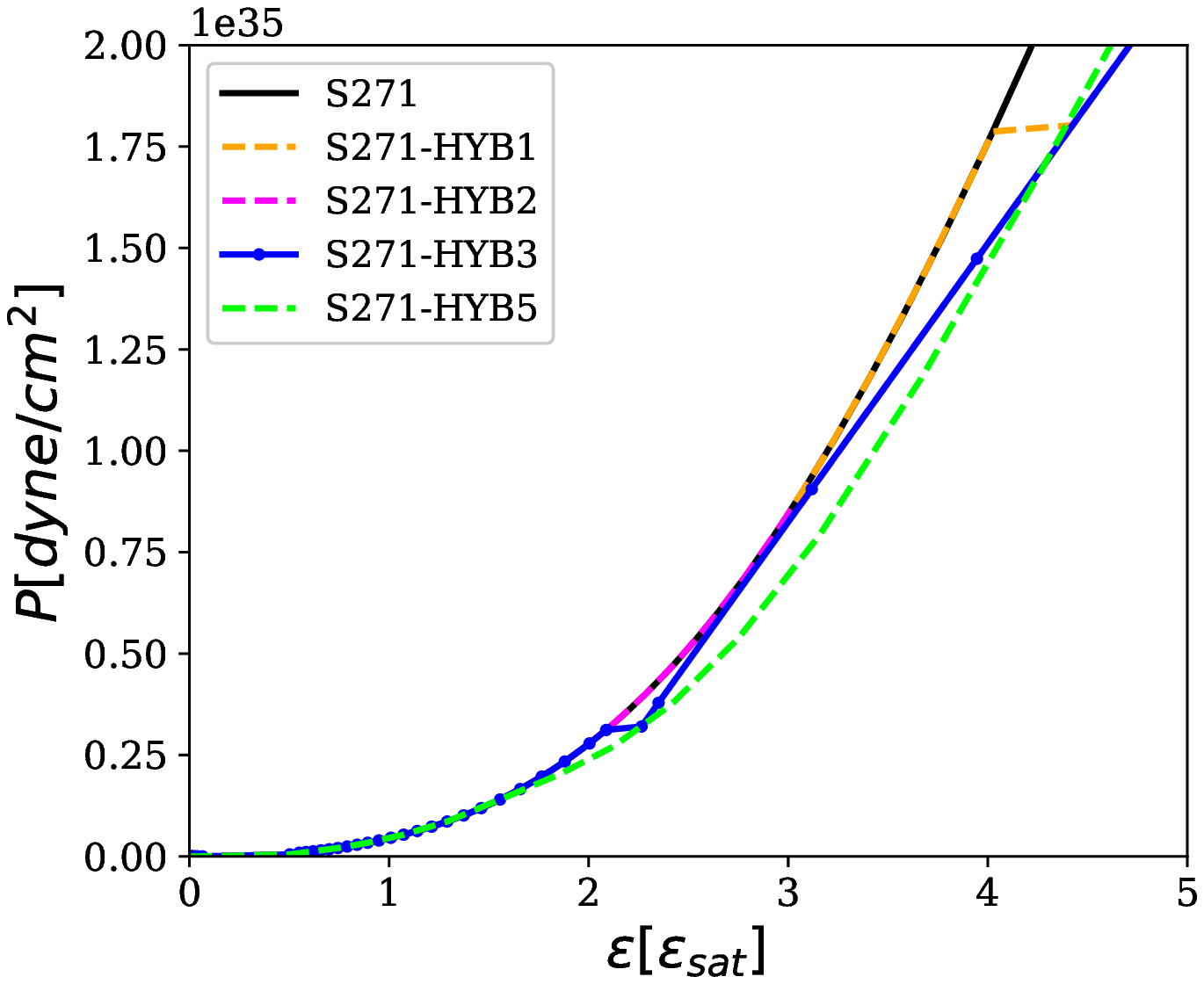}
\caption{(a) The hadronic DD2 EoS and its corresponding hybrid EoS with maxwell construction is shown. (b) The S271 EoS and its corresponding hybrid EoS with Maxwell construction and Gibbs constriction are shown. The deviation from nuclear matter to quark matter is evident in hybrid EoS, in maxwell construction the density jump $\epsilon_{N} \rightarrow \epsilon_{Q}$ occurs at transition, whereas in Gibbs construction slope of hybrid EoS changes at transition.}
\label{eos-plot}
\end{figure}

To model an NS or HS, one needs to solve the general relativistic structure equations with an equation of state (EoS) which describes the matter composition from core to crust. For pure hadronic stars, we employ DD2 \citep{PhysRevC.81.015803} and S271v2 \citep{PhysRevC.66.055803} EoS to describe the hadronic matter, and for HSs containing quark region, we employ suitable constructed hybrid EoS wherein modified MIT bag model with suitable $B_{eff}$ and $a_{4}$ parameter describes the quark matter having up, down, and strange quarks \citep{PhysRevD.9.3471, Alford:2006vz, Weissenborn_2011}. The crust is described by Baym–Pethick–Sutherland EoS \citep{Baym1971} incorporated consistently with the hadronic and hybrid EoS. The hybrid EoS is constructed using the Maxwell construction \citep{Bhattacharyya_2010}, where hadronic matter is present at low density and quark matter at high density. The critical pressure where hadronic matter ends and quark matter starts is characterized by a density jump  $\epsilon_{N} \rightarrow \epsilon_{Q}$, the transition point is ($\epsilon_{crit}$, $p_{crit}$). We carry out the study with Maxwell construction, and for comparison, the Gibbs construction is considered for a brief analysis. The Gibbs mixed-phase formalism consists of hadronic matter at low density, mixed-phase of quark, and hadron at intermediate density, followed by pure quark matter at high-density \citep{Bhattacharyya_2010, PhysRevC.87.025804}. The constructed EoS is represented in figure \ref{eos-plot}, and the details are presented in table 1. These EoS satisfy the recent constraints on mass and tidal deformability, that is $M_{max} > 2.0 M_{\odot}$, and radius of $1.4 M_{\odot}$ star lies between $13-14$ km \citep{PhysRevLett.120.172702,Rezzolla_2018}. For Maxwell construction, the deviation of hadronic EoS and hybrid EoS shows the transition point post which pure quark matter appears is assumed to occur at a few times the nuclear saturation density $\epsilon_{sat}$. For a given hadronic EoS, we have assumed three different transition points giving rise to three different hybrid EoS (figure \ref{eos-plot} (a) and (b)). The transitions occur at $2-4$ times the nuclear saturation density. 

\section{Phase transition scenario: Spin down of neutron star}
NSs are expected to attain high spin frequency at their birth after the supernova explosion. If the angular momentum is assumed to be conserved in the collapse event, most NS's birth frequency can be close to the theoretical maximum Kepler frequency $\omega_{k}$, which is the maximum frequency with which a NS can rotate without shedding mass. In realistic cases, the newly born stars are usually some fraction of the keplerian frequency. The star slows down during its lifetime due to magnetic braking, gravitational wave emission, etc. For a rotation frequency $\omega$ attained, it will remain stable only if its mass satisfies the relation: $ M(\omega) \leq M_{max} (\omega) \leq M_{max} (\omega_{k})$, where $M_{max} (\omega)$ is maximum mass which can be supported for a particular value of rotation frequency $\omega$ and $M(\omega)$ denotes the gravitational mass. Any NSs rapidly rotating at birth on a spin down during lifetime attains $M (\omega) > M_{max} (\omega)$ will become unstable at that particular rotation frequency and will collapse into a black hole (BH) or shed the mass to remain stable \citep{Falcke2014,mallick2022semi}. Another possible transition of NS due to its spin-down can occur when the central density shoots above the critical PT density ($\epsilon_{c} \geq \epsilon_{crit}$) \citep{PhysRevLett.79.1603, Staff_2006}. When such a stage is reached, nuclear matter to quark matter conversion happens via quark-hadron PT resulting in the formation of quark core. Some NSs may have central densities exceeding the critical PT density even at birth. The appearance of 3-flavor quark matter is likely to happen via 2-step PT, deconfinement in strong interaction time-scale, and weak decay in about a few seconds \citep{PhysRevC.74.065804}.  

\begin{figure}
\center
\includegraphics[height=3.6in,width=5.4in]{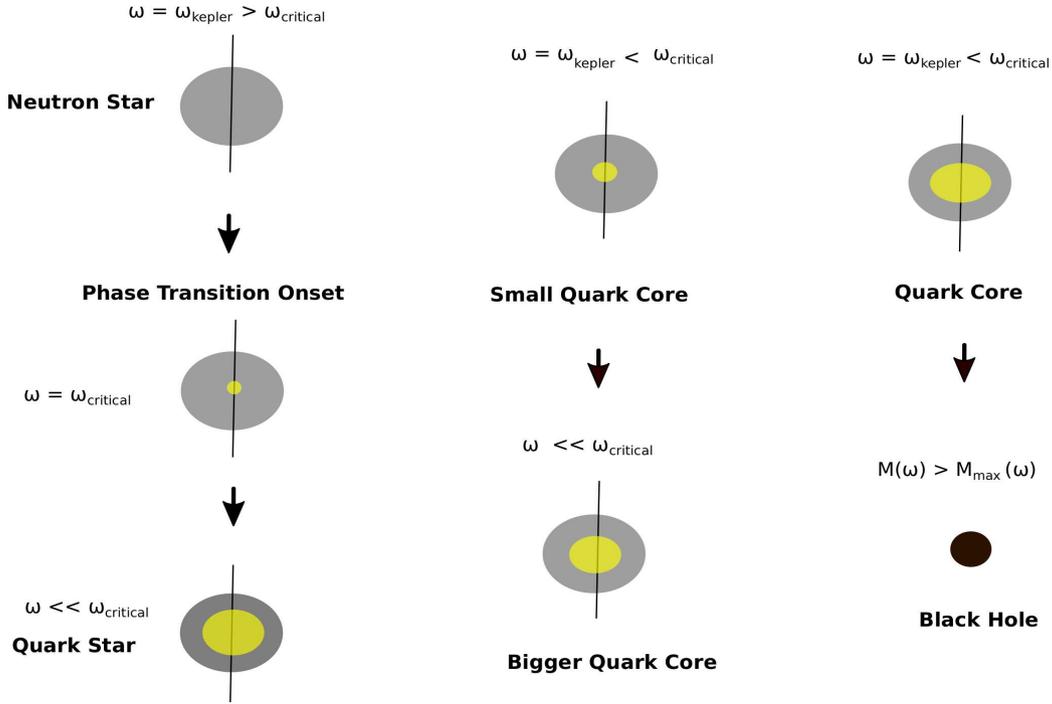}
\caption{Different evolution scenarios from spin-down: NS to QS transition, NS to BH transition, and quark core growth during evolution for a by-birth QS is depicted in the figure.}
\label{scenarios}
\end{figure}

\begin{figure}
\center
\includegraphics[height=2.8in,width=3.4in]{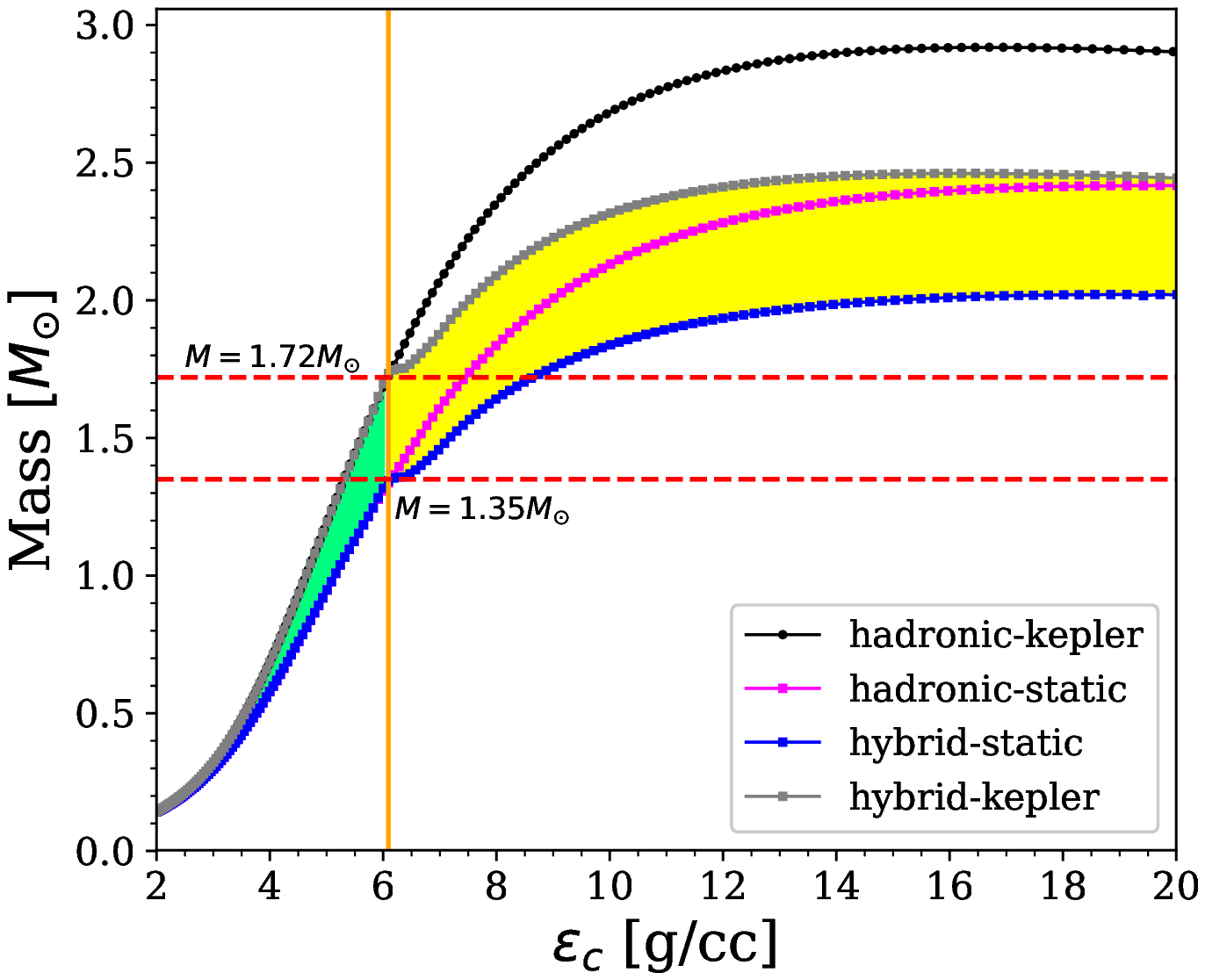}
\includegraphics[height=2.8in,width=3.4in]{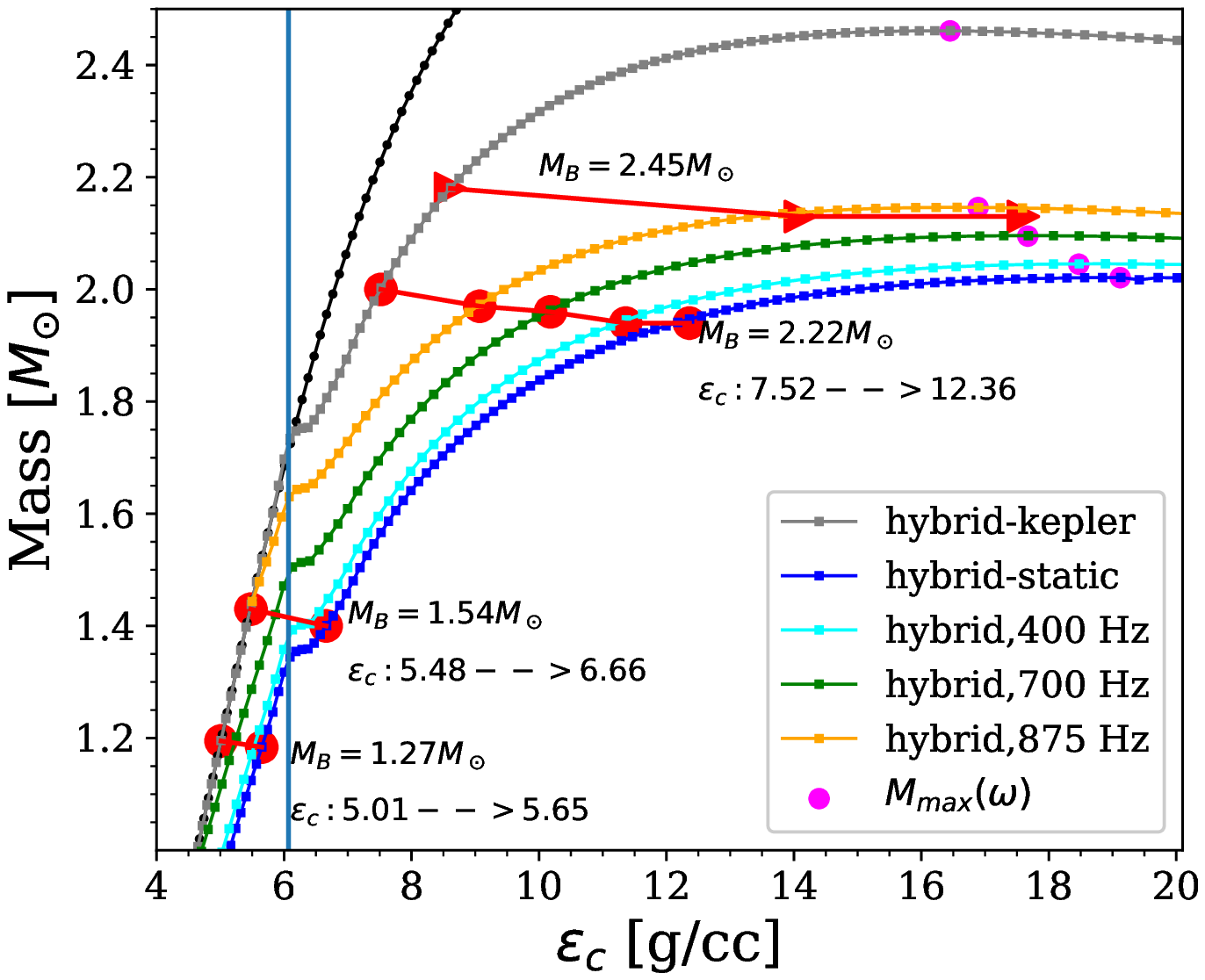}
\caption{The NS sequence for DD2 EoS and DD2-HYB2 hybrid EoS is shown for $\nu =\nu_{k}, 0$ Hz case. (b)The sequences are for additional values of rotational velocity. The evolutionary tracks of four stars of different baryonic/gravitational mass are shown.}
\label{sequence-plot}
\end{figure}

We aim to study the NS to BH transition and the NS to hybrid/quark star transition, as schematically represented in the fig. \ref{scenarios}. For simplicity, we assume that at birth, the frequency is close to keplerian frequency, that is $\omega_{birth} \approx \omega_{k}$. We construct rotating equilibrium models of NSs with different values of central density for a fixed angular velocity; this gives the mass-radius sequence. We use Rapidly Rotating Neutron Star (RNS) code for the stellar models and sequences which is based on the Komatsu, Eriguchi, and Hachisu scheme \citep{Stergioulas1995}. We can obtain different mass-radius sequence by varying the rotational velocity of the star ($\omega$). In the present work, we have different sequences for $\omega=\omega_{k}$, $\omega=0$, and intermediate rotations. In fig. \ref{sequence-plot} (a), the NS and HS sequence are shown for static models and models with keplerian frequency for DD2 and DD2-HYB2 EoS. The sequences for intermediate rotation lie between these two frequency values. The vertical line represents the PT density $\epsilon_{PT}=2.60 \epsilon_{sat}$. It can be seen that at low density, the sequences of NS and HS overlap and starts to diverge from the PT point. Stars with central density $ \geq \epsilon_{PT}$ having mass $ \geq 1.72 M_{\odot}$ are likely to attain quark core at birth (even though rotating rapidly with Kepler frequency), whereas stars with mass $ < 1.32 M_{\odot}$ will be having no quark content at birth. Stars having mass $ < 1.32 M_{\odot}$ will not contain a quark core at their birth, and on spin-down (even close to $\omega =0$), their central density will not increase beyond the critical density; hence in such stars, the quark content will not appear in their evolutionary history. The stars which are in the mass range $1.32-1.72 M_{\odot}$ on sufficient spin-down in their lifetime may reach central densities enough to initiate PT.

Since the baryonic mass of a NS remains constant during its spin-down, the baryonic mass can be used as an identifier to trace the evolution of a NS as it slows down and jumps across different NS sequences. In fig \ref{sequence-plot} (b) we track the $1.45 M_{\odot}$ star's evolution, with its baryonic mass $M_{B}=1.54$ as identifier. The keplerian frequency comes out to be $878$ Hz, and we find its evolutionary track along with sequences for $\nu= 875,700,400,0$ Hz. The central density is initially $5.48 \times 10^{14} g/cc$ which increases and crosses $\epsilon_{PT}$ about 610 Hz; the quark core first starts to appear at this frequency. Further spin-down leads to a rise in central density, indicating the growth of quark matter core inside the NS. Thus a pure hadronic star has converted to a HS in its lifetime. Next, we show the evolutionary track of $2.0 M_{\odot}$ star using its baryonic mass $2.22 M_{\odot}$ as an identifier. The star at birth has a central density above $\epsilon_{PT}$ thus, quark matter is already present in its core, and as the star slows down we find that its central density rises significantly, thereby making the quark core grow in mass and volume.

The spin-down evolutionary tracks also suggest that the gravitational mass of NSs decreases with a decrease in frequency. The change in gravitational mass can be associated with the energy released in the PT process. The evolutionary track of $1.2 M_{\odot}$ star shows that the central density never shoots beyond $\epsilon_{PT}$, and it remains a pure hadronic star in its lifetime. The maximum mass of static case is $M_{max}(\omega=0)$, and the corresponding baryonic mass is $M_{B, max}(0)=2.327 M_{\odot}$, suggests that all NSs born with a baryonic mass more than this value will eventually become unstable due to spin-down, some may get unstable while at high rotation itself whereas others will require spin-down. When such stars slow down relatively, they are prone to gravitational collapse to a BH. However, the frequency at which they collapse into a BH depends on how massive the present $M_B$ of the star is from $M_{B, max}(0)$. The evolutionary track of $M(\omega_{k})=2.18 M_{\odot}$ star whose $M_{B}=2.45 M_{\odot}$ is shown in fig. \ref{sequence-plot} (b). As it slows down to $\nu = 875 Hz$, it remains stable, whereas on further slowing down before reaching $\nu=700 Hz$, it becomes unstable and collapses into a BH. All stars having $ M_{B} (\omega) \geq 2.327 M_{\odot}$ or $M (\omega) \geq 2.02 M_{\odot}$ (maximum mass of stable non-rotating star) will collapse to a BH in their lifetime due to spin-down.

\begin{figure}
\center
\includegraphics[height=2.8in,width=3.4in]{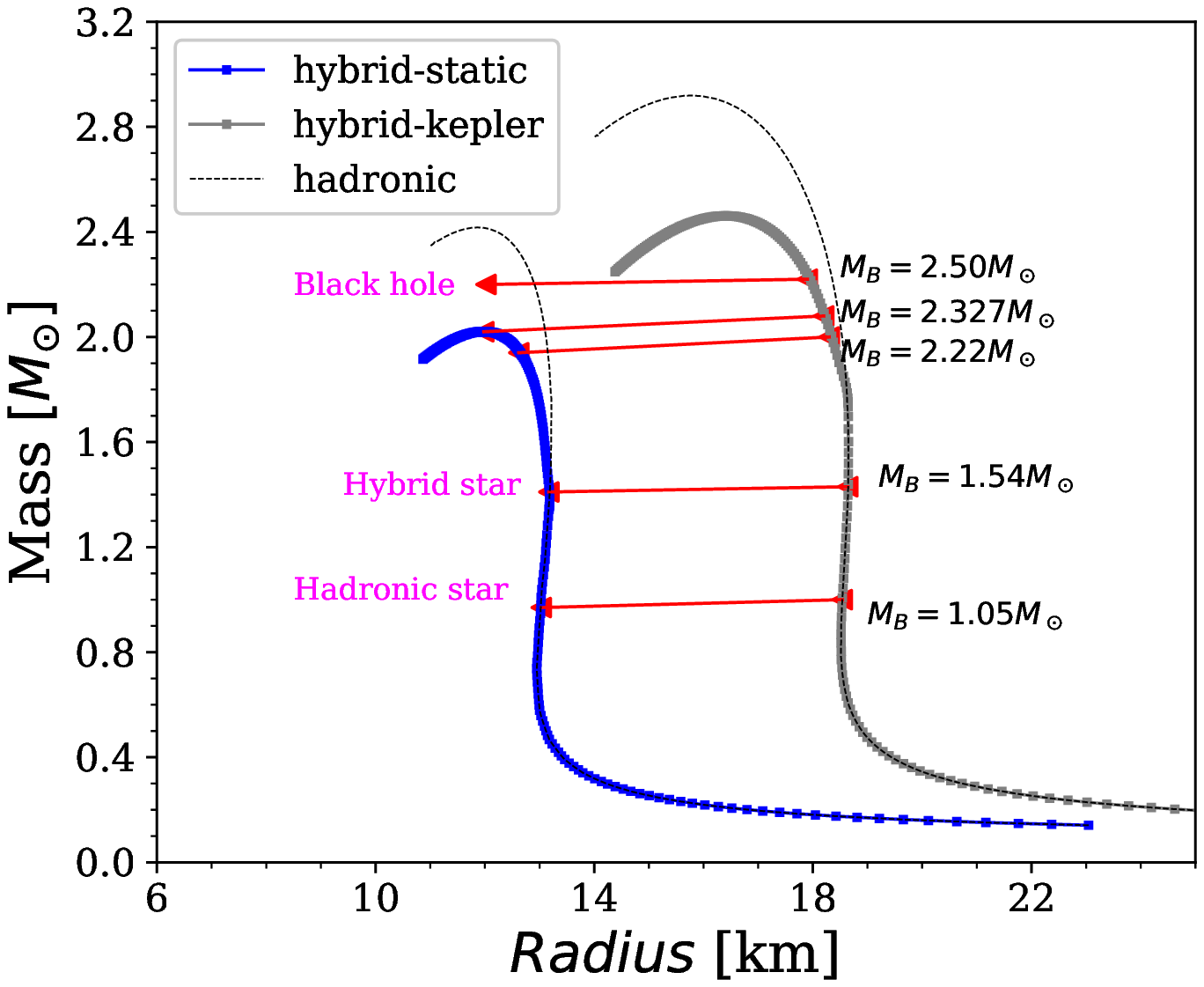}
\includegraphics[height=2.8in,width=3.4in]{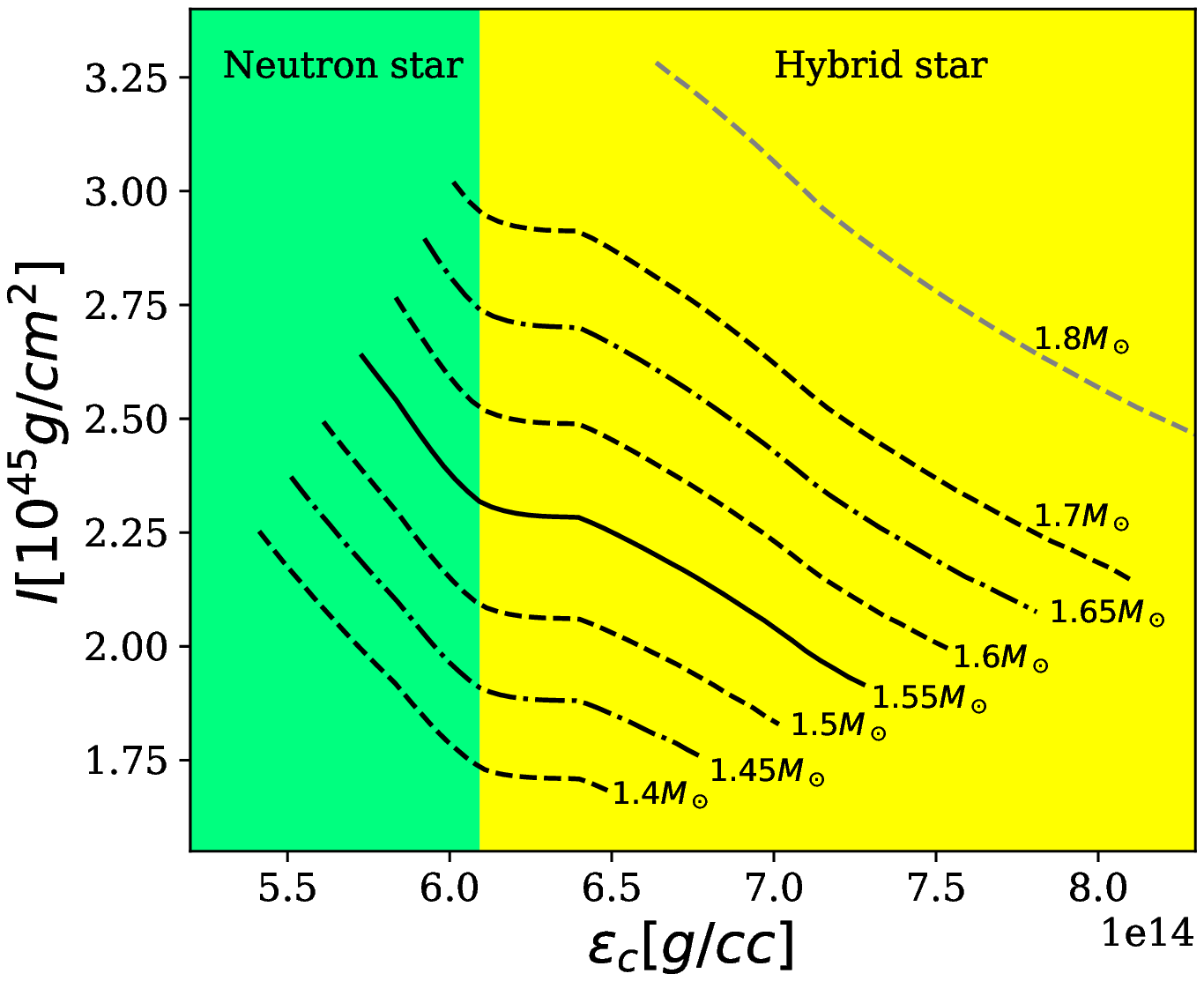}
\caption{(a)The mass versus radius curve for DD2-HYB2 EoS is shown for $\nu =\nu_{k}, 0$ Hz case. The evolutionary tracks of 5 stars of different baryonic/gravitational mass are shown. The 4 possible scenarios of BH formation during evolution, HS/QS getting compact during evolution without collapsing to BH, NS star becoming HS during spin-down evolution in its lifetime, and NS (hadronic star) remaining purely hadronic but getting compact are shown. (b) Moment of inertia evolution of SDIQC candidates for DD2-HYB EoS ($M=1.32M_{\odot}-1.72M_{\odot}$) is shown.}
\label{mr-plot}
\end{figure}

The evolutionary track in the mass versus radius diagram is shown in fig \ref{mr-plot} (a). It is evident that, as the NS slows down from kepelerian frequency, it becomes more compact. All the stars with $M_{B} (\omega) \leq 2.327 M_{\odot}$ on slowing down become more compact where both their gravitational mass and radius decrease. Therefore, stars that are born even with a quark core upon slowing down to static stars can remain stable (only their quark core grows). Stars above this mass, upon slowing down will eventually collapse into a BH, as is shown in the figure.
 
Stars rotating with keplerian frequency are highly deformed in an elliptic shape which continuously evolves during the spin-down process, leading to slower rotation, larger compactness, and lower ellipticity. The moment of inertia evolves with time due to the change in shape and also due to the composition change when the quark matter appears. The evolution of the moment of inertia is presented in fig. \ref{mr-plot} (b). The moment of inertia changes its slope substantially post the appearance of the quark content at the star's interior. This is absent in stars that never attain quark core during evolution and interestingly also absent in the stars which possess quark core at birth itself. Although their evolution will result in quark core growth, no substantial change in the moment of inertia is observed. Only in stars that attain quark core during evolution the substantial change in moment of inertia is seen. 

Next, we calculate the quark core size inside the star as it spins down. For a star of radius R ($R_{e}$ equatorial radius and $R_{p}$ polar radius for axisymmetric star) and assuming that the quark core present inside the star extends up to $R^{core}$ ($R^{core}_{e}$ equatorial radius and $R^{core}_{p}$ polar radius of quark core) which is obtained from the critical PT density  $\epsilon(r,\theta)=\epsilon_{crit}$. The quark core mass fraction (QCMF) is given by $\frac{M_{core}}{M_{star}}$, which is suitable to quantify quark core growth in the interior of the NS. The $M_{core}$ can be obtained by the formula \citep{Nozawa1998},

\begin{equation}
M_{core}= 2 \pi \int_{0}^{R^{core}} \int_{0}^{\pi} r^2 \sin \theta dr d\theta dr \Biggl[ e^{2 \alpha + \beta} 
\left\{ {(\varepsilon + p)(1+v^2) \over 1-v^2} + 2 p \right\} \nonumber  + 2 r \sin \theta \omega e^{\beta} {(\varepsilon + p) v \over 1-v^2} \Biggr]
\end{equation}

\begin{figure}
\center
\includegraphics[height=2.8in,width=3.4in]{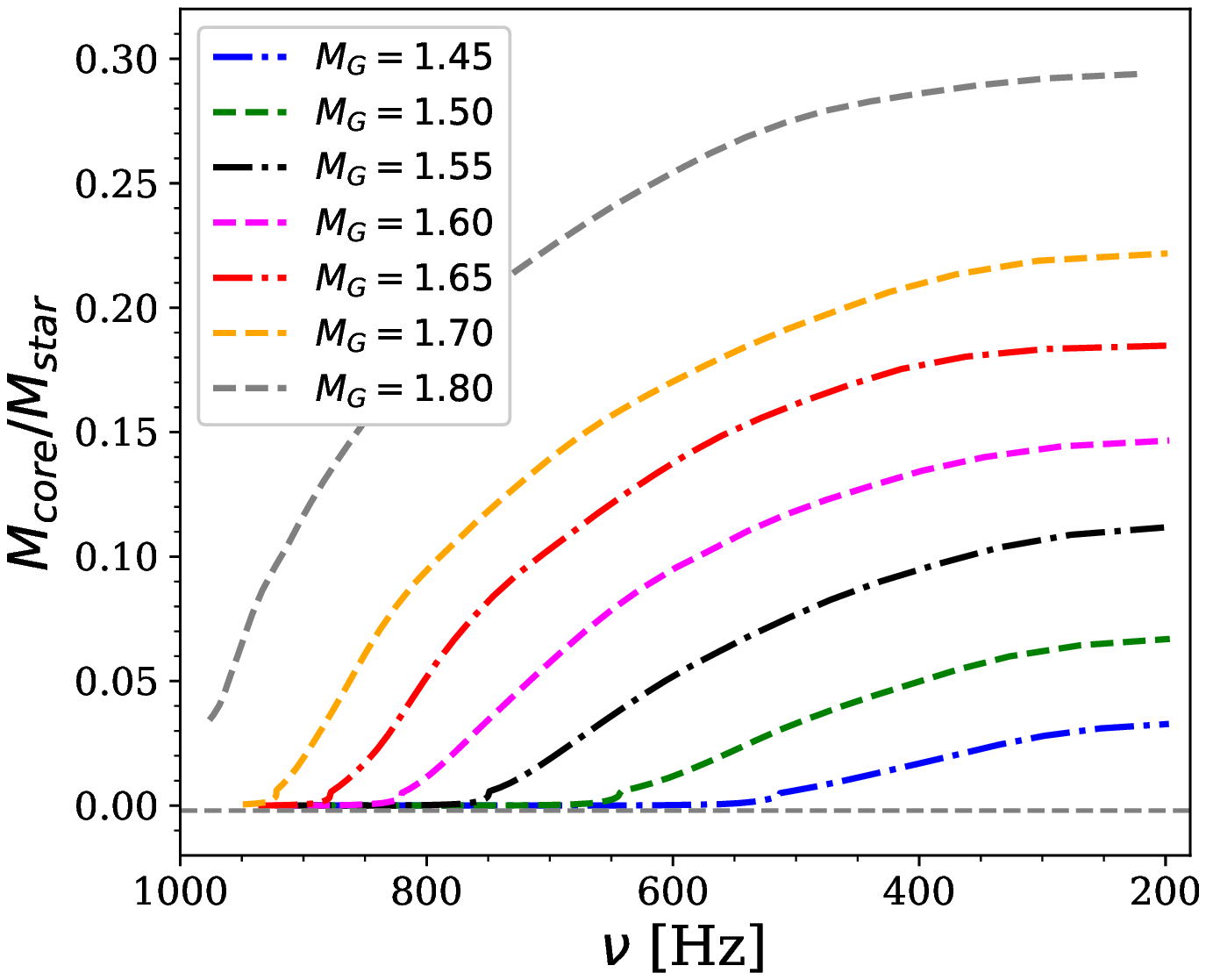}
\includegraphics[height=2.8in,width=3.4in]{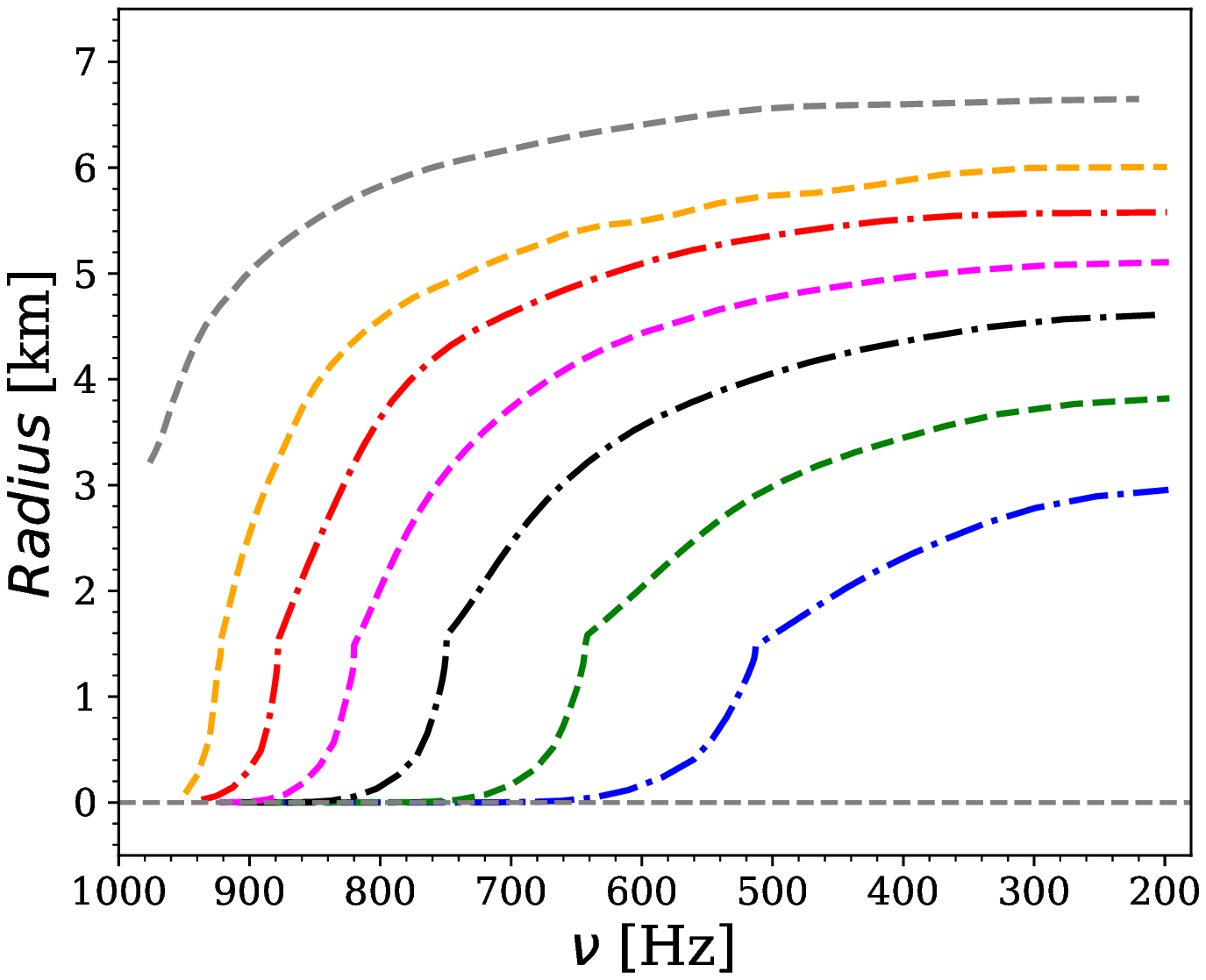}
\caption{The quark core growth during spin down evolution is shown in the figure. (a) Mass fraction of quark core versus rotation frequency (b) Quark core radius versus rotation frequency. In the manuscript both $M$ and $M_G$ are equivalent and refers to gravitational mass.}
\label{core-variation}
\end{figure}

In fig. \ref{core-variation} we show the evolution of mass fraction for stars with $M_{G}=1.4-1.70 M_{\odot}$ which attain quark content during evolution. For $M_{G}=1.6 M_{\odot}$ at Kepler frequency (assuming at birth), there is no quark core in the interior at birth. As it slows down, the quark core appears when its frequency is between 900-800 Hz and grows as it slows down further. At a small frequency (slow rotation), the QCMF is about $0.15$ and extends to a $5$ km radius in the equatorial direction. The star's radius meanwhile shrinks during its spin-down evolution. For $M_{G}= 1.8 M_{\odot}$ at Kepler frequency (assuming at birth), there is a quark core of QCMF $0.035$ of $ \approx 3$ km radii in the interior, which grows during evolution. At a very slowly rotating stage, the QCMF is about $0.30$ and extends to $\approx 7$ km. The calculations carried out are EoS dependent; however, the dynamics of spin-down induced evolution of NSs will effectively remain similar irrespective of the chosen EoS. The mass range of stars that collapses to a BH due to spin down and the mass range where quark core formation occurs at birth or during evolution will differ, but a similar overall trend is expected. We carry out the study with different EoS, and the main results are listed in table 1, the mass ranges for a NS and QS formation at birth, and also spin-down induced quark core candidates (SDIQC) and spin-down induced BH candidates (SDIBH) for evolution stage are mentioned. The SDIQC range depends on the location of critical PT density, a lower critical density suggests less massive stars as SDIQC candidates, whereas a higher critical PT will result in massive stars as SDIQC candidates. For DD2-HYB1 and DD2-HYB3, the candidate range is $1.03 M_{\odot}-1.31M_{\odot}$ and $1.72 M_{\odot}-2.22M_{\odot}$ respectively. In the Gibbs construction EoS S271-HYB4 and S271-HYB, the mixed-phase of hadronic and 3-flavor quark matter occurs at quite low density hence most intermediate and massive NSs are born containing a mixed phase in their core. The SDIQC candidate window for the two EoS (Gibbs) is $0.26-0.29 M_{\odot}$ and $0.68-0.83 M_{\odot}$ respectively. Therefore, only NSs having such small masses are born as pure hadronic stars and attains a mixed-phase quark core in its evolution. However, the existence of such low-mass NSs is almost improbable \citep{Strobel2001, Lattimer2004Sci}. The mixed-phase cores of intermediate and massive NSs will grow on a spin-down; however, the considered EoS never reaches densities of pure quark matter; hence a pure quark core is not possible for this EoS. However, there can be EoS for which pure quark cores can be reached via spin-down evolution, and such EoS will have a NS to mixed-phase star transition and additionally a mixed-phase to the pure-quark PT, leading to two distinct subclasses of QSs, the mixed-phase QSs, and pure QSs \citep{mallick2022semi}.

\begin{table}
\centering
\caption{Different hybrid EoS used in the present work are tabulated below. The mass range of NS and HS formation at birth, also spin-down induced quark core (SDIQC) candidates, and spin-down induced BH (SDIBH) candidates during the evolution stage are listed. The critical PT density for Gibbs construction EoS is mixed PT density are mentioned; however, the pure quark PT density is not listed here as it occurs at $ > 2 \times 10^{15} g/cc$. For Gibbs EoS SDIQC candidates represents spin-down induced mixed-phase quark star candidates.}
		\begin{tabular}{@{\extracolsep{2pt}}ccccccccc}
		\hline
		 	EoS &  $B_{eff},a_{4}$ & Construction & Critical PT density & \multicolumn{2}{c}{Birth Stage} & \multicolumn{2}{c}{Evolution stage} & \\ 
		  \cline{5-6}  \cline{7-8}
	& &  & $\epsilon_{crit}$ (g/cc))  &	  Neutron Star  &  Hybrid Star  & SDIQC & SDIBH  \\  	
			\hline
	       DD2-HYB1 &  (145, 0.520)& Maxwell   &   $5.2 \times 10^{14}$  & $< 1.03 M_{\odot}$ &  $ > 1.31 M_{\odot} $ & $1.03-1.31 M_{\odot}$ & $ > 2.01 M_{\odot}$  \\
		   DD2-HYB2 &  (145, 0.526)& Maxwell   &  $6.0 \times 10^{14}$  & $< 1.32M_{\odot}$ &  $ > 1.72 M_{\odot}$ & $1.32-1.72 M_{\odot}$ & $ > 2.02 M_{\odot}$  \\		   
		   DD2-HYB3 &  (145, 0.530)& Maxwell   &  $7.3 \times 10^{14}$  & $< 1.72 M_{\odot}$ &  $ > 2.22 M_{\odot}$ & $1.72-2.22 M_{\odot}$ & $ > 2.05 M_{\odot}$  \\
		   S271-HYB1 & (145, 0.508)& Maxwell   &  $4.7 \times 10^{14}$   & $< 0.94 M_{\odot} $ &  $ > 1.18 M_{\odot}$ & $0.94-1.18 M_{\odot}$ & $ > 2.02  M_{\odot}$  \\                  
		   S271-HYB2 & (145, 0.516)& Maxwell   &  $6.9 \times 10^{14}$  & $< 1.50  M_{\odot}$ &  $ > 1.90 M_{\odot}$ & $1.50-1.90 M_{\odot}$ & $ > 2.03 M_{\odot}$  \\
		   S271-HYB3 & (145, 0.518)& Maxwell   &  $9.2 \times 10^{14}$ & $< 1.92  M_{\odot}$ &  $ > 2.40 M_{\odot}$ & $1.92-2.40 M_{\odot}$ & $> 2.03 M_{\odot}$  \\
		   
           S271-HYB4 &  (145, 0.500)&Gibbs   &  $2.20 \times 10^{14}$ & $< 0.26 M_{\odot}$ & $ > 0.29 M_{\odot}$ & $0.26-0.29 M_{\odot}$ & $ > 2.25 M_{\odot}$ \\     
		   S271-HYB5 &  (162, 0.600)&Gibbs   &  $3.95 \times 10^{14}$ & $< 0.68 M_{\odot}$ & $> 0.83 M_{\odot}$ &  $0.68-0.83 M_{\odot}$ & $ > 2.09 M_{\odot}$ \\ 
			\hline
		\end{tabular}
\end{table}

\begin{figure}
\center
\includegraphics[height=2.8in,width=3.4in]{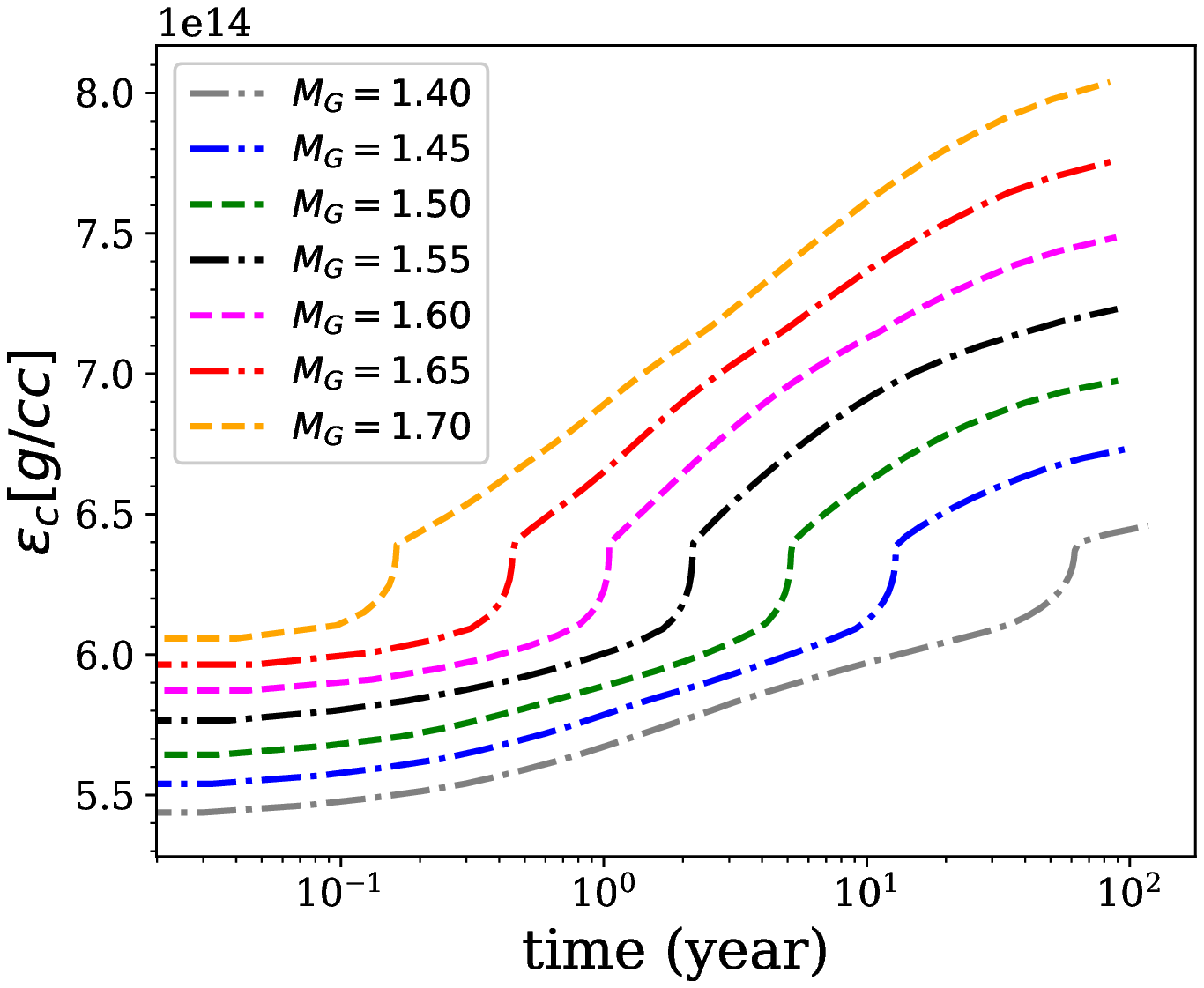}
\includegraphics[height=2.8in,width=3.4in]{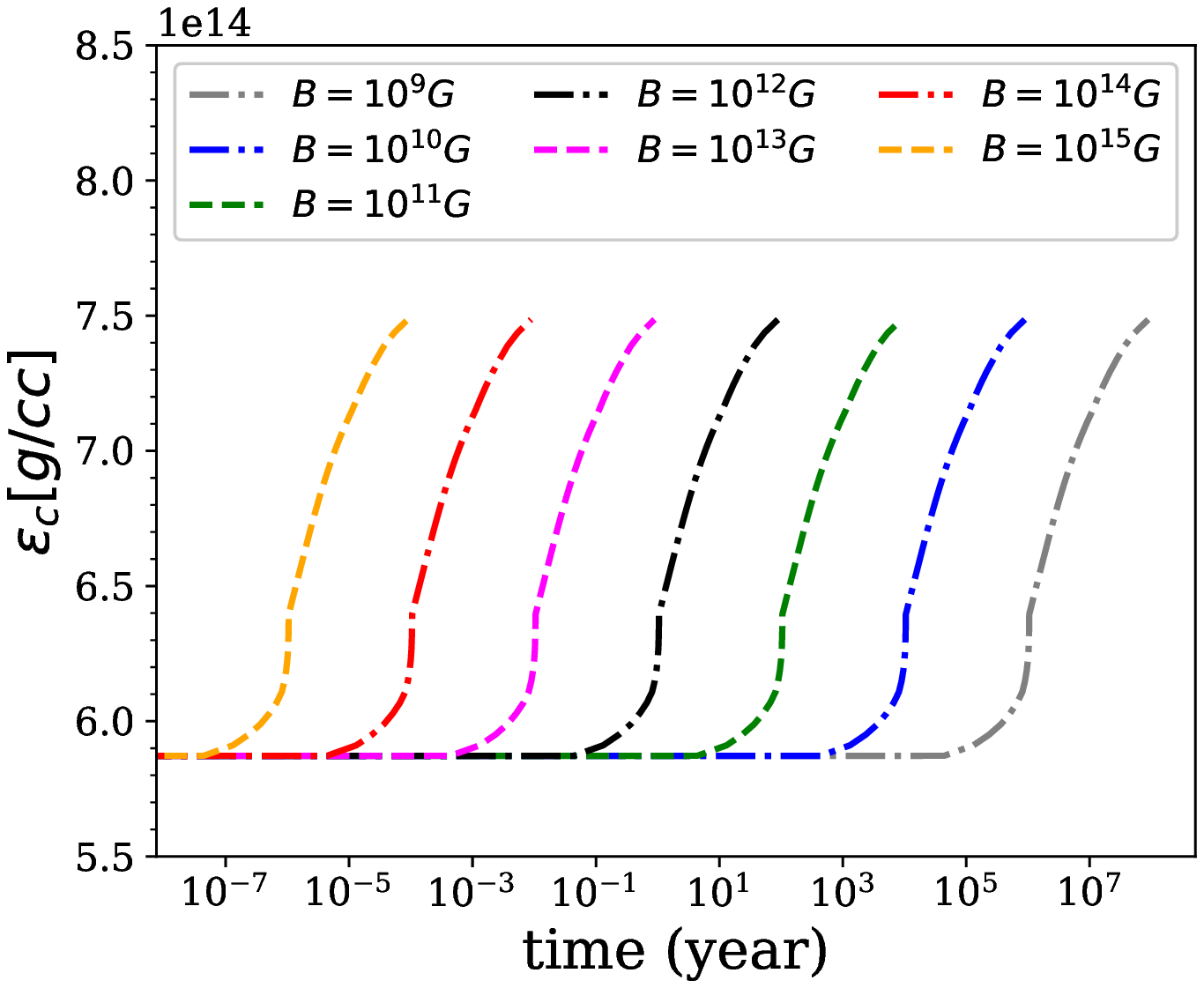}
\caption{(a) Evolution of central density ($\epsilon_{c}$) for NSs of different mass during spin down for a surface magnetic field of $10^{12}G$ is depicted. (b) Evolution of central density ($\epsilon_{c}$) for $1.6 M_{\odot}$ NS for different values of surface magnetic field $10^{9}-10^{15}$ G.}
\label{rho-variation}
\end{figure}

The spin-down of a NS can be described using the vacuum dipole model. Assuming a magnetic field of the form of a magnetic dipole inclined at an angle $\beta$ with the rotation axis of the star, the NS losses energy by emitting electromagnetic radiation due to the rotation. Under this model, the NS's period change over time and is determined by \citep{Malov2001},

\begin{equation}
P=\sqrt{P_{0}^{2}+2 A t},
\end{equation}

\begin{equation}
\dot{P}=\frac{A}{\sqrt{P_{0}^{2}+2At}},
\end{equation}

where $P_{0}$ is period at birth, $t$ is the age (time), and $A=\frac{8 \pi^{2} B_{s}^{2} R^{6}}{3 c^{3} I}$. The surface magnetic field is $B_{s}=B_{0} \sin(\beta)$, the asymptotic value $B_{0}$ and $\beta$ is taken to remain constant over the time. For a sequence of NS models generated at constant baryonic mass with frequencies ranging from Kepler frequency to static case (slow-rotating case), the time to sweep through the consecutive models ($\Delta t$) via spin-down can be estimated using the formula, 

\begin{equation}
\Delta t= t_{i+1}-t_{i} =\frac{P^{2}(t_{i+1})-P^{2}(t_{i})}{2 A_{i}}
\end{equation}

where $i$ is the index corresponding to a NS configuration at time $t_{i}$ having spin period $P(t_{i})$ with $A_{i}$ as function of its stellar properties. $i=0$ corresponds to the birth stage, and higher $i$'s correspond to the evolution stage. The time evolution of central density is presented in fig. \ref{rho-variation}(a) for stars with surface magnetic field $10^{12}$ G. Plots for different stars having mass in the range $ M=1.32-1.72M_{\odot}$ (which are spin down induced quark core candidates) for DD2-HYB2 EoS is shown in the figure. The time where central density suddenly shoots corresponds to the time when the quark matter first appears at the core. The massive stars reach critical PT density faster than less massive ones, and the time difference is about $10^3$ years between $M= 1.4 M_{\odot}$ and $M= 1.7 M_{\odot}$ stars. Since the spin-down depends on the value of the surface magnetic field, the PT onset strongly depends on the values of the surface magnetic field. The time evolution of central density is shown in \ref{rho-variation}(b) for $M= 1.6 M_{\odot}$ for surface field strengths in the range $10^{9}-10^{15}$ G. The PT critical density is attained sooner in NSs of the high magnetic field due to their faster spin-down. For a star with mass $1.6 M_{\odot}$ having the surface magnetic field in the range $10^{15-12}$ G,  magnetic braking can slow them enough to a stage where the quark core starts to develop within a year from its birth. For lower magnetic fields in the range $10^{9-10}$ G, this can take between $10-10^4$ years. The time scale will be further longer for less massive NSs. 
 
 However, the magnetic fields don't remain constant during the evolution of NSs, and it decays over the years. A simple relation which represents decaying surface magnetic field is, $B=B_{0} e^{-t/t_{D}}$ where $t_{D}$ is the decay time-scale. By incorporating the changing magnetic field, the period of evolution under the dipole model gets modified as,
 
 \begin{equation}
 P=\sqrt{P_{0}^{2}+2 A t_{D} \left( 1- e^{-t/t_{D}}\right)}.
\end{equation} 
 
This can be employed to obtain the better time estimates for beginning of PT era. The time to sweep through consecutive models in the sequence of models is given by,

\begin{equation}
\Delta t=t_{i+1}-t_{i}=-t_{D}\ln{\left( 1- \frac{P^{2}(t_{i+1})-P^{2}(t_{i+1})}{2 A_{i} t_{D}} \right)}
\end{equation}

\begin{figure}
\center
\includegraphics[height=2.8in,width=3.4in]{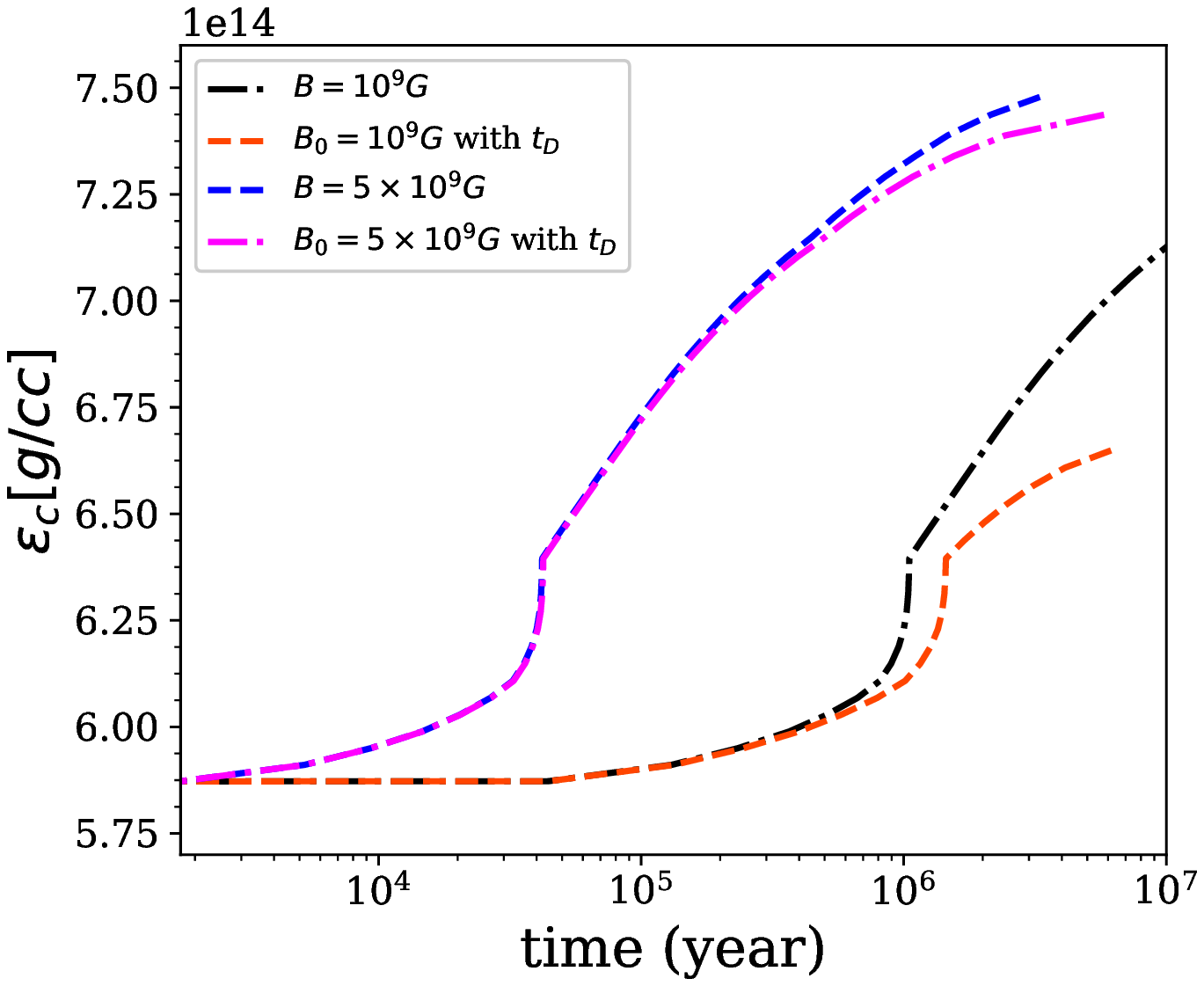}
\includegraphics[height=2.8in,width=3.4in]{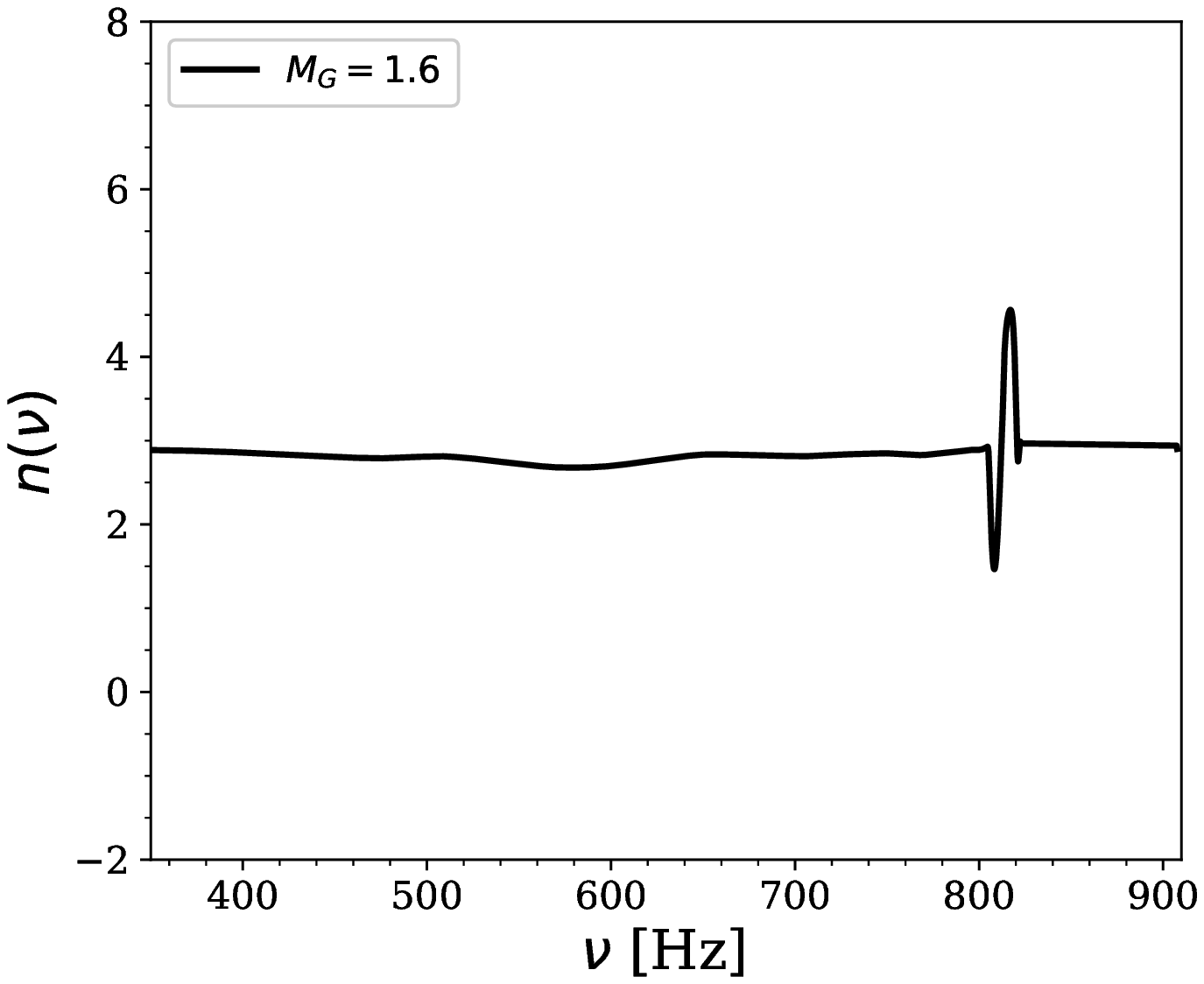}
\caption{(a) Central density evolution during spin-down due to magnetic braking caused by the constant magnetic field and decaying magnetic field for $1.6 M_{\odot}$ is shown. (b) Magnetic breaking index as a function of rotation frequency is presented and it indicates an anomalous behavior between $900-800$ Hz at PT onset.}
\label{npulsar}
\end{figure}

For NSs, the characteristic time of magnetic field decay comes out to be $10^{6-9}$ years \citep{Malov2001}. The magnetic field at the crust decays due to ohmic losses in this time scale. The decay of magnetic field occurs substantially only when the age of NSs approaches $10^{6-9}$ years, NSs with age few orders lesser than this time scale will not experience significant field decay or its impact on evolution. The evolution of central density for a star of mass $1.6 M_{\odot}$ for decaying magnetic field case for a surface field $5 \times 10^{9} G$ and $10^{9} G$   at birth is shown in fig. \ref{npulsar} (a). Decaying magnetic fields comes into play before the onset of PT and slows down the onset of the PT process for $10^{9} G$, whereas for $5 \times 10^9 G$, it comes into play much later after the onset of PT, hence it will not impact the onset and first instance of core appearance but will only impact the quark core growth. The $t_{D}=10^{6-7}$ years will not influence the PT onset for $1.6 M_{\odot}$ which has a surface magnetic field of $10^{12-15}G$, as PT happens much earlier before the decay of magnetic field comes into play. Canonical NS of $1.4 M_{\odot}$ whose evolution time-scale is in $10^{6}$ years for a birth magnetic field of strength $10^{11} G$ will undergo delayed PT onset due to decaying magnetic field. The time for the onset of PT also varies with the characteristic time; a smaller $t_{D}$ will decay the magnetic field heavily and can lead to a significantly long time for onset.

\section{Implications on stellar properties and emissions}
The onset of PT during spin-down results in a change in the moment of inertia and its derivatives, which can leave substantial markers on quantities depending on the moment of inertia. One such quantity which is also of observational significance is the braking index. We adopt the formula for magnetic braking in terms of the moment of inertia, its derivative, and angular velocity \citep{spyrou2002spin}. The formula is consistent for any rotation rate, which is a refinement of the expression provided by Glendenning \citep{PhysRevLett.79.1603}. The magnetic braking index is given by,

\begin{equation}
n(\omega) =\alpha-1 ~-~ \frac{2I' \omega+ + I'' \omega^2}{I+I'\omega},
\label{nOmega}
\end{equation}

where $I'=dI/d\omega$ and $\alpha$ govern the total mass-energy lost as electromagnetic or gravitational radiation, the typical value for magnetic braking is $\alpha=4$, which gives $n=3$ in the absence of a moment of inertia change. The change in moment of inertia at the onset of PT results in a change in the magnetic-braking index. For $1.6 M_{\odot}$ the braking index during spin-down evolution undergoes a sharp transition near 800 Hz, represented in (fig \ref{npulsar} (b)). Observationally, in measured pulsar data, any anomalous behavior exhibited by braking index will indicate the onset of PT and the critical frequency and critical density where diversion from hadronic phase to quark phase can be extracted. High precision measurement of spin frequency and its derivatives is necessary for accurate values of magnetic braking. Newly born and young pulsars having rapid rotation, low timing noise, or glitches are ideal sources where the onset of PT can occur. In the case of old, slow-rotating, or low magnetic field pulsars, the spin-period change ($\dot{\omega}$, $\ddot{\omega}$) measurement itself may take many years. The NSs which acquire density above critical density at birth and those NSs which never reach critical central density by spin-down in their lifetime will never exhibit an anomalous change in magnetic braking. This anomalous behavior is confined only to those NSs that, on a spin-down, sweep through critical central density.

The PT can also excite the stellar quasinormal modes (QNMs), which could lead to gravitational wave emission. The star's QNMs are of different types, namely fundamental modes (f-mode), pressure modes (p-mode), gravity modes (g-mode), toroidal modes (r-mode), and space-time modes (w-mode) based on the restoring force which brings the star to equilibrium during the pulsation/ oscillation phase. The quasinormal modes contain a real part and imaginary part, corresponding to the mode's oscillation frequency and the mode's damping time scale. In the present work, we focus on the f-mode and its corresponding gravitational wave as its frequency is $ \sim 1-2$ kHz. The GWs produced due to any QNM is time-varying signals which decay exponentially depending on the mode's damping time scale and is described by \citep{marranghello2002phase, PhysRevC.83.045802},

\begin{equation}
\label{gw}
h(t)=h_{o} \exp \left(-\frac{t}{\tau_{f}} \right)  \sin(\omega_{f} t),
\end{equation}

where $h_{0}$ is amplitude given by \citep{pacheco2001potential},

\begin{equation}
\label{strain-amp}
h_{0} = \frac{4}{\omega_{0} D} \sqrt{\frac{\Delta E}{\tau_{f}}},
\end{equation}

where $w_{f}$ is mode frequency, $\tau_{f}$ is damping time, $\Delta E$ is the energy associated with it and $D$ is the distance of the source. Although the baryonic mass remains conserved during spin-down induced PT the gravitational mass changes. The change in gravitational mass results in enormous binding energy release from the process, which is given by $\Delta E = M^{NS}-M^{QS}$. The released energy can excite the oscillation modes of the star, enhancing the gravitational wave amplitude substantially for detection. To calculate the energy emitted, we consider that the star remains in a meta-stable phase for a short period, and the gravitational binding energy is released together. The meta-stable phase may represent a stage where the pressure is equal to critical pressure, but the density is changing intermediate between the hadronic phase $\epsilon_{N}$ and quark phase $\epsilon_{Q}$ in which hadronic matter can be taken to be in a stable super-dense phase which eventually gets unstable and the density jump $\epsilon_{N} \rightarrow \epsilon_{Q}$ would occur, a possibility considered in Ref. \cite{Zdunik2008}. For ease of calculation, we assume the metastable phase at the center of the NS exists (lasts) for a duration equivalent to sweeping 0.5 Hz via spin down. After PT on-set, the time to pass through $0.5$ Hz for $1.6 M_{\odot}$ NS (birth gravitational mass) having $10^{12} G$ surface field via spin-down is about 2.3 days; during this period the gravitational mass change is $4.02 \times 10^{-5} M_{\odot}$ and the energy released comes out to be $7.19 \times 10^{49} ergs$. A major portion of this energy is likely to get converted into different emissions originating, like the GWs, neutrinos, and gamma-ray bursts, and the remaining may get used up by other dissipative processes depending on their time scales. Since the energy budget available for gravitational wave emission is uncertain, we assume $10$ percent of the available energy is converted into GWs.

The f-mode oscillation frequency scales linearly with $\sqrt{\frac{M}{R^{3}}}$. The oscillation modes of the non-rotating star can be extracted from the linearized equations in general relativity, which involves fluid and metric perturbations \citep{Thorne1967,Thorne1969,Detweiler1985,doi:10.1098/rspa.1991.0016}. Due to its complexity, usually, Cowling approximation is employed where metric perturbations are neglected, which gives values in satisfactory agreement with values obtained from the full general relativistic formulation 
\citep{Kokkotas:1999bd,PhysRevC.95.025808}. Several empirical formulas have also been obtained for ease of calculations for a wide range of EoS 
\citep{1998MNRAS.299.1059A,PhysRevD.70.124015}. The calculation of quasinormal modes or rotating NSs involves sophistication and is at an early stage; with this regard we calculate the f-mode oscillation frequency and damping time using the following relations \citep{yim2022},

\begin{equation}
    \omega^{2}_{f}= \frac{4 GM}{5 R^3} , \hspace{2cm} \tau_{f} =\frac{625}{32}\frac{c^5}{G^3} \frac{R^4}{M^3}.
\end{equation}

\begin{figure}
\center
\includegraphics[height=2.6in,width=3.4in]{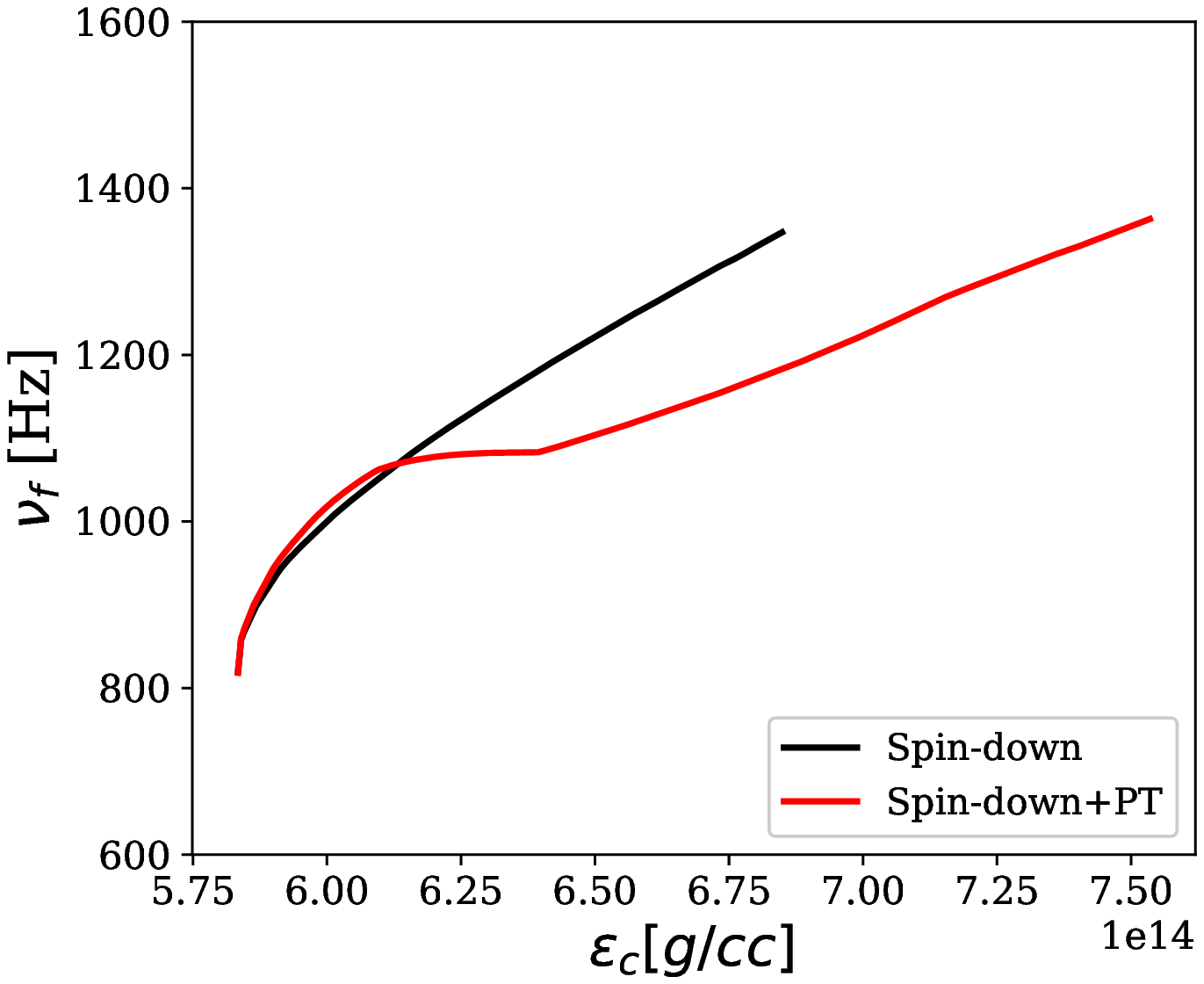}
\includegraphics[height=2.8in,width=3.4in]{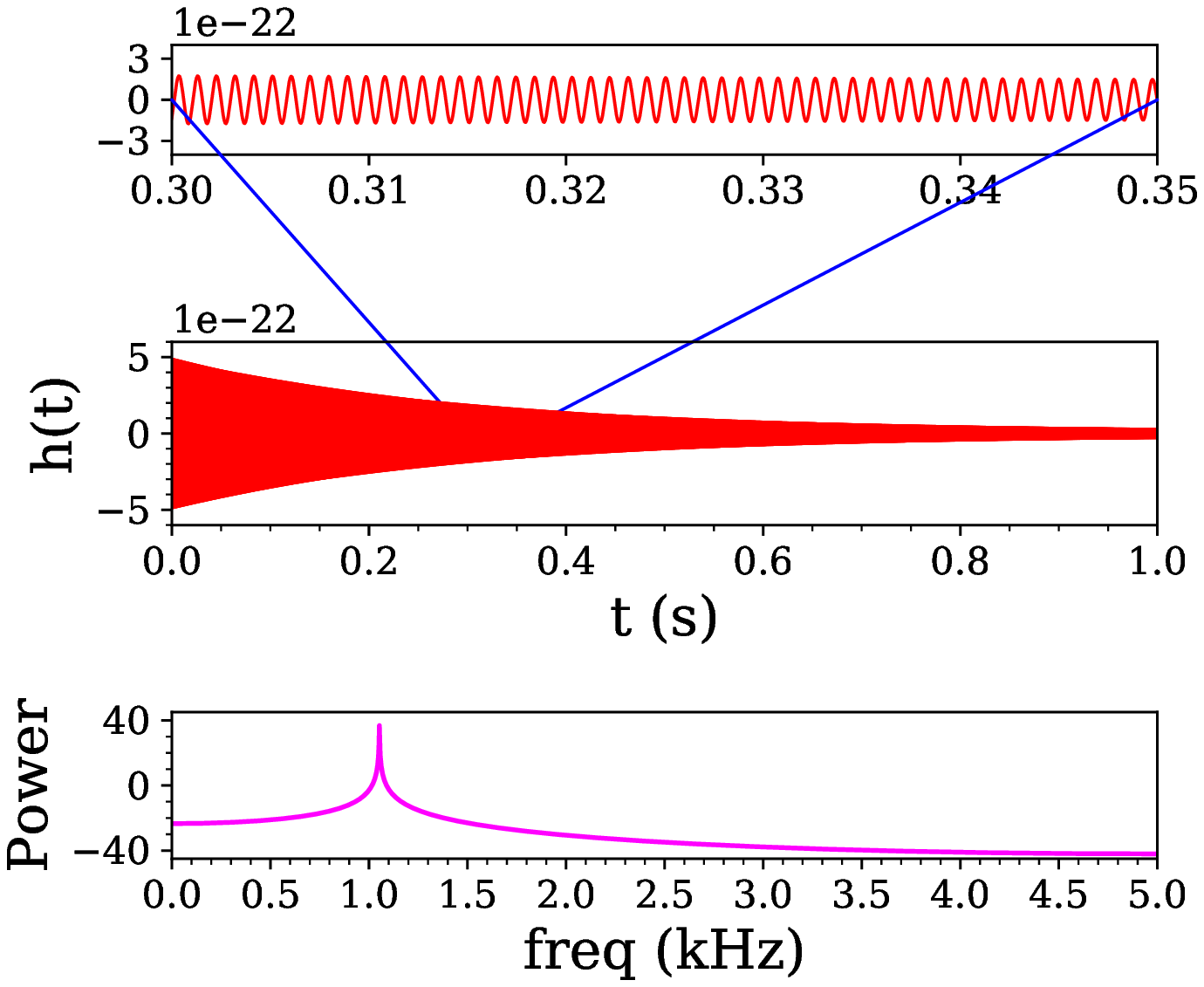}
\caption{(a) The evolution of f-mode's oscillation frequency during spin down caused by the magnetic braking and for a spin down induced PT case for $1.6 M_{\odot}$'s evolution is compared in the figure. (b) The gravitational wave strain due to f-mode excited by energy released and periodogram (scaled) is shown at onset of PT.}
\label{fmode}
\end{figure}

The QNMs depend on the macroscopic properties and composition of NSs. The f-mode frequency evolution during pure spin-down and spin-down associated with PT is presented for $1.6 M_{\odot}$ in fig. \ref{fmode} (a). The f-mode frequency's flattening near the $6 \times 10^{14} g/cm^3$ indicates the onset of PT, post which f-mode's frequency increases with density. It can be seen that the f-mode frequency evolution due to spin-down associated with PT is different from that of the pure spin-down case. The deviation of f-mode frequency can serve as an indicator for PT. The f-modes frequency comes out to be $1052.8$ Hz, and the damping time-scale is $\sim 0.29$ s. The gravitational wave strain for a NS at 100 kpcs undergoing spin-down induced PT is shown in fig. \ref{fmode} (b), with the assumption that the fraction of energy dissipated by GWs is $10$ percent of the total energy released. The signal is shown for a $1$-second duration, its amplitude is $\approx 10^{-22}$. The decaying amplitude is evident due to the f-modes damping time scales. A portion of the signal is magnified, which brings out the sinusoidal nature of the signal, and the periodogram is also shown, which has a peak at $1052.8 Hz$, which is the f-mode frequency after PT. The NSs that are invisible as radio-pulsars due to their presence in the graveyard region of the $p-\dot{p}$ diagram or due to the emission-axis orientation towards the earth or their large distance may be observed via QNMs excited GWs at the onset of PT and during stages of quark core growth. With upcoming third-generation detectors, Einstein Telescope and Cosmic Explorer, which possess sensitivity higher than the sensitivity of Advanced LIGO, cleaner extraction of burst signal will be achievable for GW strains of amplitude $10^{-24}$ in the kHz frequency band. The GW signal amplitude would be lower if the energy budget available for GW emission is less than $10$ percent as assumed, a $10^{-4}$ percent of total energy available will indicate strain amplitude to be $10^{-26}$ for a 100 kpc source distance, and in such the f-mode driven GW emissions from galactic sources ($\approx 10-20$ kpc) near to earth could reach us with amplitude suitable for detection. The f-mode's damping time is proportional to the spin period of the NS ($ \propto P^{4}$), and the oscillation frequency is  $\propto \sqrt{M}$ and  $\propto \sqrt {R^{3}}$, hence from the obtained GW signal's oscillatory and damping behavior the stellar properties can be estimated. The fast-rotating stars will have smaller damping times and higher gravitational wave strain, evident from Eqn. \ref{strain-amp}. Thus f-mode oscillations offer new windows to measure the properties of NS and HS in the gravitational wave regime. Also, the information regarding true EoS obeyed in nature can be obtained by studying these modes from spinning down NSs, which in turn can shed light on the location of critical PT density at the intermediate and high-density range. Another interesting oscillation mode, is the discontinuity g-mode which is associated with density discontinuity alone and appears only in NS/HS possessing density discontinuity in the interior. The g-mode frequency can be obtained using, 

\begin{equation}
 \omega^{2}_{g} \simeq l(l+1)\,{\mathcal C}^2\,\frac{R-R^{core}}{R} \text{  where  }  {\mathcal C}^2 \equiv \bar{\rho}_{\rm_{core}} \,\frac{\Delta\rho/\rho_{\rm crit}}{1 + {\Delta\rho}/\rho_{\rm crit}},
\end{equation}

expressed in terms of average density of quark core $\bar \rho_{core}$ and $\Delta \rho$ is density jump which occurs at critical mass density $\rho_{crit}$ determined by EoS \citep{miniutti2003non}. The discontinuity g-mode oscillation frequency is $\propto \sqrt{R-R_{core}}$ and also $ \propto \sqrt{\bar\rho_{core}}$ in which information of core and its core radius is contained, and its typical value is $\approx 0.8-1.25$ kHz, However, this discontinuity g-mode's damping time $\tau_{g}$ is greater than $10^{4}$s \citep{zhao2022universal,miniutti2003non}, hence this mode will not play a major role as dissipating mechanisms for damping the PT excited stellar oscillations. Also, the corresponding gravitational wave strain-amplitude is given by Eqn. \ref{strain-amp} will be much lesser than the f-mode, and it will also not be a burst-type gravitational wave signal. 

Additionally, the strange matter formation via weak decays in the two-step PT could result in abundant neutrino generation, which would propagate to the surface on a millisecond or second-time scale depending on the matter temperature and its neutrino opacity \citep{BHATTACHARYYA2006195,mallick2013edr,10.1093/mnras/stab2217}, and a neutrino burst signal can be observed. The energy scale $(10^{49} ergs)$ is in the range of short GRBs \citep{DAVANZO201573, Berger2014}. Very massive NSs close to the maximum mass of the rotating mass-energy density sequence having quark core (supermassive HSs) are SDIBH candidates. In such stars, after reaching their critical BH collapse frequency ($M (\omega) > M_{max} (\omega)$), the gravitational collapse will come into play and squeeze the stellar matter inside the event horizon. A big chunk of hadronic matter may undergo deconfinement at the last stage. The subsequent strange matter formation and neutrino transport to the surface will occur in $ > 1$ ms time-scale; however, since the free fall time-scale for BH and event horizon formation occurs in $ < 1$ ms \citep{Falcke2014}, any final neutrino signal produced by a rotating NS collapsing into a BH may not be visible. If the true PT critical density is about $2-3$ times the nuclear saturation density (intermediate density range), all massive NSs collapsing into a BH are HSs containing quark cores attained at birth. 

Previous calculations concerning PT-induced modes have considered catastrophic PT wherein a full-fledged quark core extends up to the maximum attainable distance and gets formed \citep{ferrari2003gravitational,marranghello2002phase, PhysRevC.83.045802,dexheimer2018phase}. There the quark matter (EoS dependent) appears instantly, and a single f-mode frequency and damping time govern the GW emission, which differs from the spin-down scenario presented in this work. The amount of quark core a star can support strongly depends on the rotation frequency (which governs the stellar central density and thus quark composition); hence gradual PT is likely to occur in rotating NSs which represents a different picture from a catastrophic PT event. We have carried out energy released and f-mode-driven gravitational-wave estimates only at the onset of PT when the first quark core appears. After the onset of PT during further slowing down, the quark phase extends to new regions within the star; it can happen gradually or in multiple leaps as it passes through the meta-stable hadronic phase accompanied by a small density jump ($\epsilon_{N} \rightarrow \epsilon_{Q}$) as it engulfs new regions of the star. The corresponding emissions in the form of GWs and neutrinos would occur, which could be persistent or multiple short-duration transient signals. Alongside magnetic braking driving spin down, the r-mode instability can also come into play, leading to substantially speeding up of the spin-down \citep{Staff_2012, Staff_2006}. 
%Also, the spin-down is associated with back bending phenomena first predicted by \citep{PhysRevLett.79.1603} and is exhibited depending on the EoS (hadronic, 2-flavor, 3-flavor matter) and also on rotation frequency at which deconfinement PT occurs \citep{chubarian2000deconfinement,spyrou2002spin}. 
Also, deconfinement PT can lead to enhancement of magnetic field in highly magnetic NSs \citep{Franzon2016} and may be the reason for recurring large glitches \citep{spyrou2002spin}. 

\section{Summary and Conclusion}

To summarize, in the present work, we have studied the spin-down induced PT in cold, isolated NSs. As the star slows down after birth, its central density rises, and once it crosses the critical density for PT a small quark core appears at the center of the star. As the star further slows down, the quark core grows with time. However, there are four possible scenarios of the spin-down evolution of an NS. Less massive stars whose central density never reaches the critical density do not ever achieve a quark core in their lifetime. Intermediate mass stars are likely to have a quark core in their life after they slow down to such a frequency that at the center, the critical density of PT is reached. Massive stars contain a quark core from their birth, and the core grows as they slow down, but they remain HSs throughout their life. Further massive stars also have a quark core from birth, but as they slow down, they collapse into a BH.

Next, the structural change in the star is further studied as the star slows down after birth. As the star slows down, it tries to restore its spherical shape from the initial spheroidal shape. It leads to a change in the moment of inertia of the star. It is shown that the evolution of the slope of the moment of inertia of stars that achieve a quark core in their lifetime after birth is different from stars that either does not achieve a quark core in their lifetime or already possess a quark core from birth. It is also seen that as the star slows down, its central energy density rises. The rise is discontinuous at the point when the quark core first appears in the star and which has a definite signal in the braking index of such stars. The slow down of the star is attributed to the rotational energy drain of the star through the magnetic poles in the form of electromagnetic waves. If such is the case, then stars that achieve a quark core in their lifetime only show a discontinuity in the braking index at the frequency when the quark core first appears at their center.

Not only a change in the braking index, but the PT also happens as the quark core takes birth and as it grows. The PT is accompanied by a release in energy and thus excites the f-mode oscillation of the star and leads to GW emission. As the star slows down by about a Hz in a few days, it releases about $10^{49}$ ergs of energy. Even if a small fraction of this energy goes into GW emission, then for a star at $100$ Kpc, the amplitude of the gravitational wave is in the range of present operating detectors. The frequency is also around the $1$ kHz range. Therefore, spin-down induced PT in cold, isolated NS can have detectable signatures upon dedicated, focused observation. The only drawback of such detection is that most of the detectable signature is associated with the first seeding of the quark core of isolated stars in their long lifetime. However, the possibility of subsequent signatures (glitches, GW signals from excited QNMs, neutrinos, etc.) could arrive in the evolution stage after the onset of PT if PT further doesn't proceed gradually and consists of subsequent leaps due to meta-stable phase and density jump occurring in the interior in new regions where quark matter is getting formed for the first time as quark core grows. 

The seeding of the quark core at a particular frequency depends on the EoS and mass of the initial NS. All isolated pulsars are rotation-powered pulsars, and they release a large amount of energy with rotational deceleration. A part of the energy is consumed in accelerating particles in their magnetosphere, which results in radio emissions. The pulsars which are undergoing spin down and PT together will have more energy budget for different emission processes, which may be persistent or transient depending on gradual PT (without leaps) or PT with leaps. Persistent or multiple transient signals would imply that NSs undergoing PT can be studied via f-modes just like the radio pulsars and would enable extracting a wealth of information about EoS and PT critical density. Thus, spin-driven PT in a cold, isolated NS is an interesting quark star formation scenario that leads to different implications on stellar properties and emissions. With the coming of proposed ground and space-based gravitational wave detectors and multi-messenger astronomy efforts, many of the theoretical predictions related to the NSs and HSs physics can be tested, paving the way towards understanding the high-density matter and compact stars.

\section*{Acknowledgements}
RP would like to acknowledge the financial support in the form of the DST-INSPIRE Fellowship from the Department of Science and Technology, Govt. of India. RM and RP would also like to thank IISER Bhopal for providing all the research and infrastructure facilities.
\bibliographystyle{mnras}
\bibliography{references}

\label{lastpage}
\end{document}